# THE CASE FOR A PUBLICLY AVAILABLE, WELL-INSTRUMENTED GBT OPERATING AT 20-115 GHZ

*J. Bally[1], G. Blake[2], A. Bolatto[3], C. Casey[4], S. Church[5], J. di Francesco[6], P. Goldsmith[2,7], A. Goodman[8], A. Harris[3], J. Jackson[9,10], A. Leroy[11], F. Lockman[12], A. Lovell[13], A. Marscher[9], D. Marrone[14], B. Mason[15], T. Mroczkowski[16], Y. Shirley[14], & M. Yun[17]*

A well-instrumented Green Bank Telescope (GBT) operating at high frequency represents a unique scientific resource for the US community. As a filled-aperture, 100m-diameter telescope, the GBT is ideally suited to fast mapping of extended, low surface brightness emission with excellent instantaneous frequency coverage. This capability makes the GBT a key facility for a range of cutting edge science, only possible with observations at 20-115 GHz:

- Large-scale, < 0.1 parsec-resolution mapping of interstellar filaments in molecular species that trace dense gas, to determine the filament turbulent and thermal pressure, mass accretion rate, and the dynamical time scale of star formation.
- Sensitive observations that map the fraction of dense gas and the distribution of gas volume densities across entire galaxies.
- Mapping of the Sunyaev-Zeldovich Effect over entire galaxy clusters with high sensitivity and complete flux recovery down to the sub-arcminute scales necessary to determine the internal pressure and column structure of the intracluster gas.
- Spectroscopic mapping and planetary radar imaging of rapidly evolving Solar System targets to determine gas production, composition, thermal structure, and solar wind-magnetic field interactions.
- Millimeter-wave VLBI observations to study the physics and masses of supermassive black holes and their associated jets.

No other observatory, in operation or planned, can perform these observations as well as a GBT equipped with focal-plane array heterodyne receivers and large 3mm bolometer cameras. For a fixed surface brightness sensitivity, the mapping speeds of such a facility are orders of magnitude faster than can be achieved with cutting edge interferometers. Meanwhile the large aperture of the GBT ensures the angular resolution needed to recover key physical information, resolving filaments, galaxies, Solar system objects, and galaxy clusters. A GBT equipped with array cameras is a much-needed complement to the Atacama Large Millimeter Array (ALMA), the Jansky Very Large Array (JVLA), a high frequency component to the Square Kilometer Array (SKA), and the planned next generation VLA (ngVLA). Interferometers are extremely powerful machines for studying compact sources, but are not designed for fast mapping over large areas with high surface brightness sensitivity.

The ability of the GBT to observe at high frequencies has been evolving rapidly over the last half decade and is reaching a mature state (see §2 and Lockman 2016). Currently the GBT is equipped with the 7-element KFPA operating at 23 GHz, the 223-element MUSTANG-2 bolometer camera operating at 90 GHz, and the 16-element ARGUS heterodyne array (in commissioning) also at 90 GHz. The current surface accuracy yields a ~40% efficiency at 3mm, so that the dish has collecting area equivalent to a 73 m-diameter dish. The amount of time available is substantial, 1,000 to 2,000 hours per year.

These capabilities can be expanded with modest effort: surface adjustment and control can increase the efficiency at mm-wave frequencies to 44-48%, and high frequency operations could be expanded to enable day-time observing. There is also a clear path forward from the current instrument suite to next generation, large arrays. The mapping speed comparison in Table 1 shows that for realistic site conditions and with current sensitivities the GBT is already faster than ALMA at large area mapping to a target surface brightness. With the improvements described in §2.a, it has the potential to become considerably faster. Ultimately, a clear path exists toward instruments that sample the full GBT field of view and combine large bandwidth and large numbers of pixels. Previous developments have leveraged partnerships between universities, the NSF, and the observatory. An efficient, publicly accessible GBT operating at high frequencies will enable such projects to proceed forward.

The future availability of the GBT to the US community is now in considerable doubt, as is its future operating at high frequencies. Indeed, a possible path for operations with reduced NSF support is to shut down all high frequency observing and concentrate on longer wavelengths. However, as reviewed by Lockman et al. (2016), the NSF portfolio review that recommended divestment occurred before any of the high frequency capabilities highlighted in this paper came into regular operation.

The incremental investment to instrument the GBT is relatively small and within the scope of the MSIP program. *We argue that rather than divesting from this exceptional resource, the US community should invest moderately to maintain GBT operations and instrument it in an optimal manner, enabling it to become an extraordinary complement to existing and future radio interferometers*. No other observatory has the capabilities of the GBT, and none has open access for US investigators to the degree offered by the GBT. Adequately instrumented, the GBT would be a pillar for 20-115 GHz science in the US and the world.

Table 1. Time to map to a target surface brightness sensitivity

| GBT 2015<br>1 Pixel<br>8" | ARGUS 2016<br>16 Pixels<br>8" | GBT 2020<br>50 Pixels<br>8" | ALMA<br>50x12m<br>3" | ALMA<br>50x12m<br>1" | ALMA<br>10x7m<br>23" | ngVLA<br>500x18m<br>1" |
|---|---|---|---|---|---|---|
| 21h | 3.3h | <1h | 19h | 1,500h | 0.5h | 17,500 h |

*Declination 0°, 1 km/s velocity resolution, Ta*, GBT 2000h/yr opacity*

*Speed to map a 3'×3' region down to $T_A^*$=20 mK at λ=3mm (Lockman et al. 2016). The ngVLA performance corresponds to the 2015 baseline design outlined in the ngVLA project web page.*

# 1. Science Cases for High Frequency at the GBT

## 1.a. Star Formation in the Milky Way: The Need for Dedicated Surveys

- **How do cluster-forming clumps evolve?**
- **What is the role of filaments?**
- **What is the distribution of high-mass star formation in the Galaxy?**
- **How can we understand star formation throughout the universe?**

High-mass stars ($M > 8\ M_{\odot}$) dominate the energy input and chemical enrichment of galaxies. Nevertheless, despite decades of effort, high-mass star formation remains poorly understood. Compared with low-mass stars, high-mass stars are rarer, their formation time-scales shorter, their clustered environments more crowded and confused, and their surrounding dust and gas much more opaque. Moreover, once they form, high-mass stars quickly disrupt their natal environments due to their large UV fields, powerful winds, and eventual supernova explosions. The two major competing theoretical models about how high-mass protostars ultimately acquire their mass, either locally via “monolithic collapse” or from afar via “competitive accretion” (see reviews by Zinnecker & Yorke 2007 and McKee & Ostriker 2007), have yet to be definitively verified or excluded by the observations.

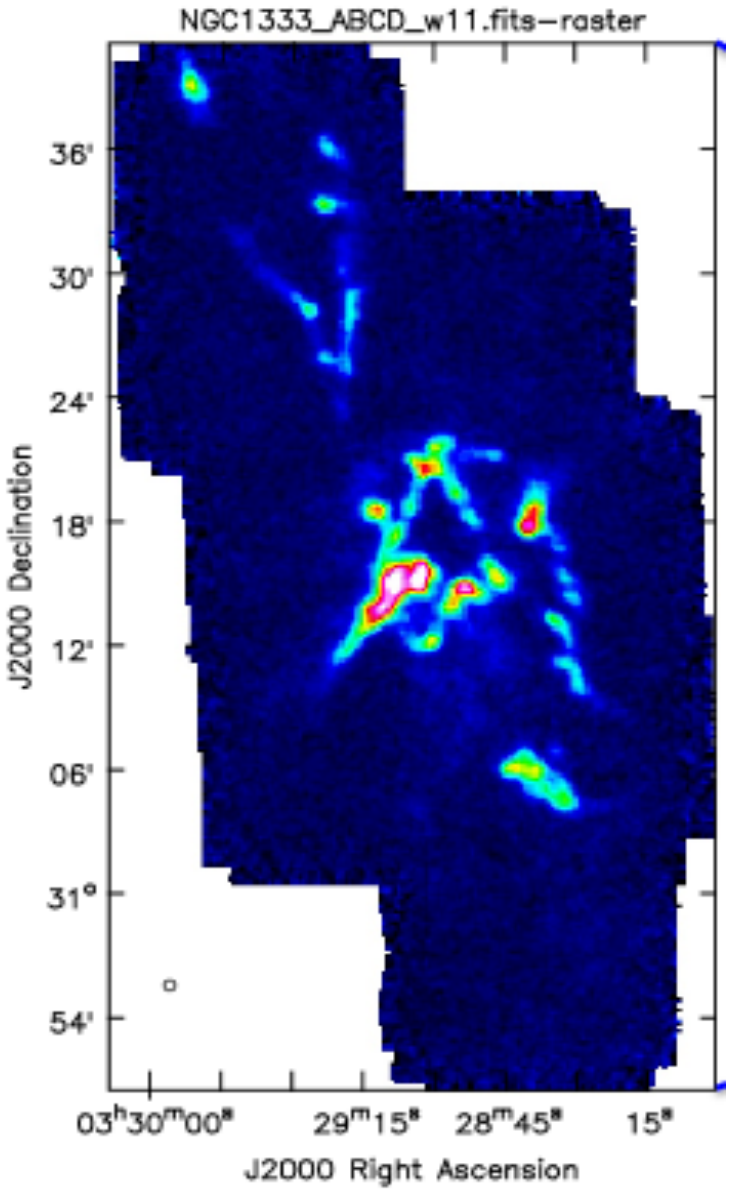

**Figure 1**. *A GBT $NH_3$ (1,1) integrated intensity image of the NGC1333 star-forming region from GAS. Note the complex filamentary/hub structure and the scale of the map needed to capture them.*

Star formation occurs within the larger context of the molecular component of the interstellar medium (Fig. 1). Observations of thermal dust continuum emission by Herschel and Planck reveal that filamentary structure is a universal aspect of the structure of molecular clouds and plays a major role in controlling the star formation that takes place within them. While optical and radio spectral line observations had previously indicated the presence of filaments, the combination of large-area and high angular resolution studies made it clear that these structures are ubiquitous, contain almost all young stars, and are the location of dense cores – the sites of incipient star formation. Filaments result from theoretical models involving colliding flows, and can also be formed via instabilities in relatively quiescent gas. Filaments are thus an ubiquitous component of massive (as well as low mass) star formation.

Millimeter and submillimeter continuum Galactic surveys (BGPS: Aguirre et al. 2011; ATLASGAL: Schuller et al. 2009 and HiGAL: Molinari et al. 2010) have now identified thousands of dense molecular clumps (~2 pc in diameter, and ~200 to 2,000 $M_{\odot}$ in mass) that are currently forming, or will eventually form, high-mass stars and star clusters. These clumps are almost invariably associated with dense filamentary infrared dark clouds (IRDCs). While ALMA has made some progress on sub-parsec scales, major scientific gains can now be realized from dedicated single-dish molecular line surveys at ~pc (typically ~1 arcmin) and larger scales unavailable to ALMA. *With the transformational* *improvements of the Green Bank Telescope's performance at high frequencies, the GBT has become the world's premier instrument for pc-scale molecular line surveys of the Milky Way.* With modest investment, new high-frequency GBT surveys can finally answer many of the open questions in star formation.

**Open Questions:** Although research into high-mass star formation has made important progress in the past few years, still a number of open questions remain:

*How do cluster-forming clumps evolve?* There remains a great detail of uncertainty about the evolution of high-mass star-forming and cluster-forming clumps. For example, their early fragmentation history and turbulent support, as well as their physical and dynamical evolution, remain poorly understood.

*What is the role of filaments?* Although recent *Spitzer* and *Herschel* results demonstrate that dense filamentary structures are common and intimately associated with high-mass star formation (e.g., Jackson et al. 2010, Molinari et al. 2010), it is unclear what role filaments play in the process. The flow of gas along filaments, their linear mass density, and the spacing of the clumps along filaments will provide important clues and direct tests to the theories of filament formation and fragmentation. Many questions about filaments remain that can be answered by observations possible with the GBT. Among them are (a) Can we determine the dominant processes in filament formation (b) Do filaments form in a stable configuration and then become unstable? (c) If the latter, is mass accretion the key factor pushing filaments over the edge to form dense cores? (d) Are there universal characteristics of filaments such as their width and is it the same in molecular tracers as in dust? (5) Are dense cores formed by instabilities in filaments, and if so, how does this impact the rate of star formation that results?

*What is the distribution of high-mass star formation in the Galaxy?* The location of high-mass star-forming regions in the Galaxy has long been an outstanding problem. Despite some successes in assessing the distribution of HII regions and ~100 clumps associated with bright masers, nevertheless it has been difficult to assign precise locations or associations with spiral arms to star-forming regions, especially the youngest ones.

*How can we understand star formation throughout the Universe?* Recent advances have allowed us to probe star-formation in nearby and distance galaxies. However, the interpretation of the molecular line and dust data relies heavily on ad hoc empirical assumptions. We need to understand star-formation in the Milky Way to understand the complex star-formation ecosystems in external galaxies.

Unfortunately, the recent large-scale continuum surveys of the Milky Way provide only partial answers, due to the following significant limitations. First, at optically thin mm, submm, and FIR wavelengths, continuum emission from every cloud along the line of sight is blended, making it difficult to isolate and separate the star-forming clumps from each other and from extraneous foreground and background material. Second, the continuum measurements cannot provide distances, and hence fail to deliver such key parameters as mass, size, Galactic location, and luminosity. Finally, the continuum surveys cannot provide critical kinematic information. Indeed, the velocity dispersion $\sigma$ is a key parameter in all theories of high-mass star formation, as it sets the turbulent pressure ($\sim\rho\sigma^2$), the mass accretion rate ($\sim\sigma^3/G$), and the dynamical time scale ($\sim R/\sigma$). In addition, the velocity fields can reveal large-scale flows; the spectral profiles can reveal collapse motions and outflows; and the linewidths can reveal the degree to which a clump is gravitationally bound ($\alpha=M_{vir}/M\sim\sigma^2R/GM$).

**The Advantages of Molecular Line Surveys:** The GBT can have a major impact on this important field as a result of the combination of angular resolution and large-scale imaging ability. Note that these frequencies give us access to the fundamental ground transitions of many species, which are easier to collisionally excite in cold molecular gas. The GBT can observe key molecular tracers of filaments (including $N_2H^+$, $HCO^+$, $C^{18}O$, $^{13}CO$) at angular resolution (in the 3mm range) of 7.5", corresponding to resolving a 0.1 pc thick filament at a distance of 2,750 pc. With the ARGUS 16 element focal plane array, large areas can be imaged in a reasonable time. As an example, a 10' x 10' region can be imaged to 0.14 K rms in 5.6 hr with velocity resolution of 0.05 $kms^{-1}$. This should give exquisite images of $C^{18}O$ emission delineating the column density distribution, together with detailed information on the kinematics of the gas (see Fig. 6 for an ARGUS commissioning map of a brighter isotopologue, $^{13}CO$, in Orion). HCN and $HCO^+$ will require significantly less time, ~ 1.5 hr. The spatial dynamic range (image size/angular resolution) is hundreds for these maps, providing the context that allows the assessment of longitudinal flows leading to core formation, and transverse flows indicative of mass accretion onto the filaments. Thus, the GBT is uniquely capable to study filaments throughout a large portion of the Milky Way, to determine their internal density profiles and kinematics, and to let us understand their past evolution as well as studying the dense cores and future star formation taking place within them. Many of the limitations of the FIR/submm/mm continuum surveys can be overcome, however, with dedicated molecular line surveys. Indeed, many recent ~1' resolution surveys such as RAMPS, GAS, and KEYSTONE at the GBT and MALT90 at Mopra are now providing some of this crucial missing information. These surveys exploit relatively bright molecular line probes of star-forming clumps at ~23 and 90 GHz, which have enormous diagnostic power. Specifically, the molecular lines of several simple molecules such as $NH_3$, HNC, HCN, $HCO^+$, and $N_2H^+$ have the following observational advantages:

**High Density Tracers:** ***The 90 GHz rotational transitions of simple molecules, as well as the 23 GHz inversion transitions of $NH_3$, are key probes of dense, star-forming clumps and cores because they require high densities for excitation ($n > 10^3$ $cm^{-3}$).*** Because star-forming clumps and cores are much denser than typical CO-emitting molecular clouds (n~300 $cm^{-3}$), the mere detection of these lines ensures that one is probing *only the dense star-forming gas*, and that all other extraneous diffuse molecular gas along the line of sight has effectively been filtered out. Thus, the 90 and 23 GHz lines probe only the dense regions of interest; *the foreground and background emission that severely contaminates the dust continuum is completely absent.* An important additional advantage is that many of these lines, especially $N_2H^+$ and $NH_3$, tend to have modest optical depths toward dense clumps. Thus, unlike CO lines, which can be compromised by extremely large optical depths, these optically thin probes can trace the internal structure throughout the clumps.

**Distances:** Without distances, it is impossible to derive luminosities, sizes, masses, and Galactic location. Although continuum data cannot provide distances, molecular line data can fill this gap simply by providing the radial velocities, which can in turn be converted into a kinematic distance with a suitable rotation curve, recently refined and calibrated by the maser parallax distances (Reid et al. 2014). The near/far kinematic distance ambiguity can be resolved with a variety of techniques, such as searching for the presence or absence of H I self-absorption (e.g., Jackson et al. 2005), H I absorption toward continuum sources (Andersen and Bania 2011), or mid-infrared star counts and color information (Foster et al. 2012). Although parallax measurements of masers provide more accurate distances, the kinematic distance method is nonetheless reliable and provides the only known technique that can supply distances to thousands of dense clumps (and exclusively so for the more typical non-maser sources) in practical amounts of observing time.

**Kinematic Information:** A major advantage of molecular line data over continuum data is the ability to supply kinematic information. In particular, the aforementioned transitions give us access to the kinematics of cores and intermediate density material in the immediate environment. In addition to the kinematic distance revealed by the systemic velocities, *the 90 and 23 GHz lines can separate overlapping, blended clumps along the line of sight* (Fig. 3) and supply velocity field information (gradients, flows along filaments), diagnostic spectral line shapes (infall, outflows, line wings, self-absorption), and the linewidths (virial parameters and dynamical time scales).

**Thermometry:** Since the 23 GHz $NH_3$ inversion transitions span a large range of excitation energies, their brightness ratios robustly indicate the *gas temperature* (see Fig. 3). Because of the lack of background emission in the $NH_3$ lines, these $NH_3$ rotational temperatures are, unlike the dust temperatures, unaffected by background or foreground sources. Gas temperatures can differ from dust temperatures; the difference ($T_{gas}$ – $T_{dust}$) establishes the gas cooling rate and the cooling time at high densities.

In addition to the thermal emission lines, the 23 – 90 GHz range also includes a number of prominent molecular masers, such as $H_2O$, SiO, and $CH_3OH$. Because these lines are so bright, they can be seen at great distances and followed up at milliarcsecond angular resolution with VLBI techniques.

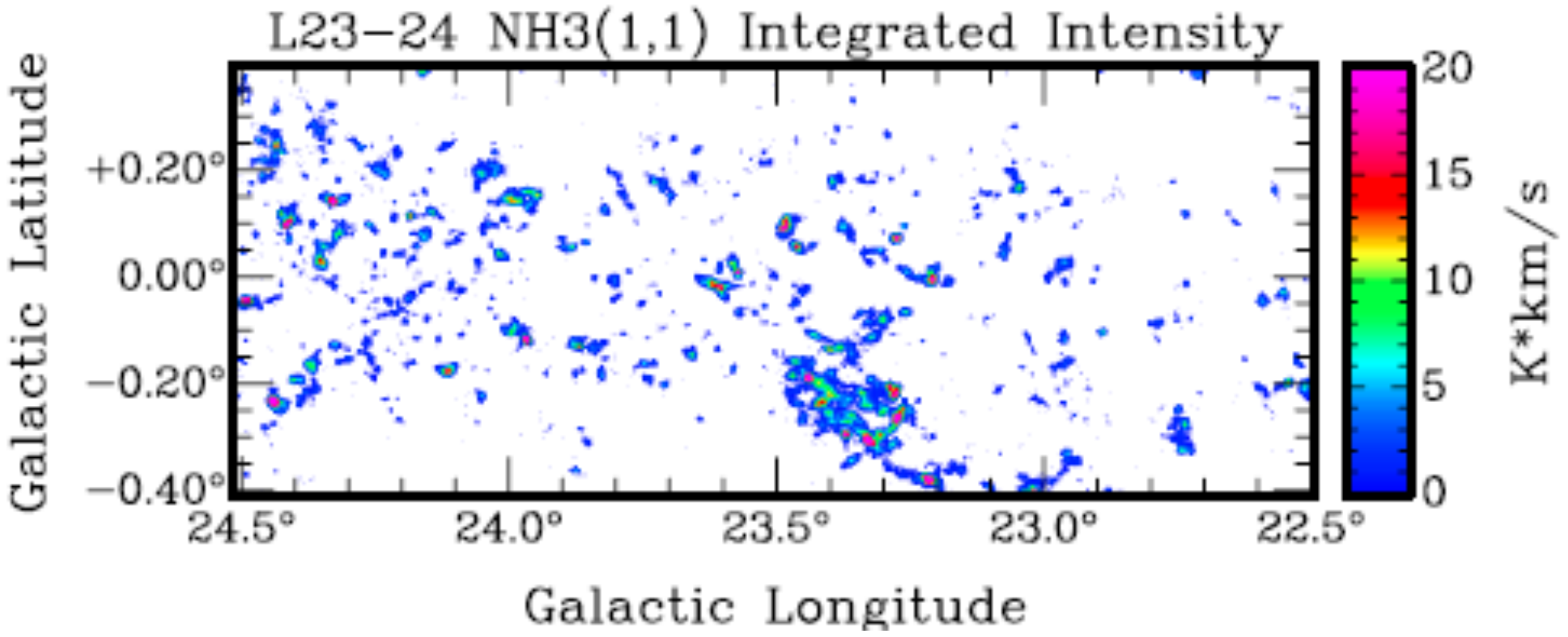

**Figure 2**. *A GBT/RAMPS NH3 (1,1) integrated intensity image of a 0.7 square degree region of the Milky Way. Hundreds of dense clumps are detected, and their temperatures, column densities, and kinematic states can be analyzed statistically.*

**Current and Future Capabilities:** Astronomers have already begun to take advantage of the GBT's extraordinary sensitivity to begin molecular line Milky Way surveys focussing on $NH_3$ thermal lines and $H_2O$ masers (GAS, RAMPS, and KEYSTONE; Figs. 1 and 2). These surveys already demonstrate the GBT's ability to perform cutting-edge surveys. Nevertheless, they are still limited by mapping speed constraints to ~10 sq degrees. Deeper, faster surveys are required to achevie the scientific goals of characterizing Milky Way's star-forming clouds.

Although GAS, RAMPS, and KEYSTONE are making important inroads into open questions as well as creating new ones, with the current sensitivity and mapping speeds only the brightest emission is easily studied. This limitation imposes an important bias: we cannot yet see the full picture of how more diffuse giant molecular clouds condense into dense clumps in the first place.

With an improved GBT and larger arrays, we can finally achieve the sensitivity to probe the fainter emission from the more diffuse gas that surrounds the dense star-forming clumps. Although the GBT is already proving itself as premier high-frequency surveying instrument, a well-instrumented GBT will achieve the transformational capability to preform molecular line Galactic plane surveys of hundreds of square degrees at the sensitivities necessary to access a full suite of physical diagnostics.

## 1.b. Dense Gas Across Other Galaxies: Imaging Faint Extended Emission

1. **How does the amount of dense and very dense gas vary across galaxies and what drives its variation?**
2. **How do chemical abundance patterns and isotopic ratios change across galaxies?**
3. **How does emission break down between photon-, X-ray, cosmic-ray, and shock-dominated environments?**
4. **How does the local star formation rate depend on the physical state of the gas across galaxies?**

The Milky Way offers the possibility for in-depth studies of the physical conditions for star formation. Nonetheless, the range of conditions present is limited. To deepen our understanding of the effect of different parameters and the mechanisms of feedback it is clearly necessary to expand the range of conditions and gather statistically meaningful samples. That is only possible through studies of nearby galaxies, practically defined as galaxies closer than z~0.05.

Star formation in molecular gas clouds represent a key step in the buildup of galaxies across cosmic time. In most types of galaxies, surveys have so far mainly been able to estimate gas masses and kinematics from CO emission. But studies of the Milky Way and other galaxies highlight a much more direct connection between dense gas and star formation. Detailed cm- and mm-wave spectroscopy offer the prospect to diagnose the *physical state* of gas across galaxies. Given sufficient sensitivity, ratios among lines in the 20-115 GHz regime can illuminate the volume density, dense gas mass fraction, temperature, and the presence of strong and weak shocks in the molecular gas of other galaxies. The relevant physical processes all play out on small scales (see previous section), which are extraordinarily challenging to directly resolve at extragalactic distances. Fortunately, density and specifically the mass of very dense gas (perhaps corresponding to the mass in the aforementioned filaments), are accessible via spectroscopic surveys that target cm- and mm-wave lines with high critical densities and contrast them with total gas tracers. Thus, mm-wave spectroscopy illuminates the small-scale physics of star formation and feedback even in systems where we cannot resolve the processes directly.

The challenge to use mm-wave spectroscopy in this way is sensitivity. Key diagnostic lines include high critical density rotational lines of HCN, HNC, CS, $HCO^+$, isotopologues like $^{13}CO$, $C^{18}O$, $C^{17}O$, photodissociation tracers such as CN, and shock tracers like SiO and HNCO. These lines are almost always fainter than the bulk-gas tracing CO lines by a factor of 10-30, with many more tracers being 100 or more times fainter than CO. This faintness has so far caused spectroscopic studies to be focused on bright starburst samples and case studies. Sensitive, fast spectroscopic mapping of wide areas - a current and future strength of the GBT at 20-115 GHz - offers the prospect to radically change this.

In the Local Group, this would allow cloud-by-cloud (~30 pc resolution) spectroscopic surveys of M31, M33, and the star forming dwarfs of the Local Group. Beyond the Local Group, the GBT can carry out spectroscopic mapping of many types of galaxies that resolve them into discrete regions on scales of a few hundred parsecs. With the sensitivity to detect many lines, the GBT has the prospect to produce a cm- and mm-wave analog to the new generation of optical IFU surveys. These spectroscopic maps would reveal how gas density, dense gas mass, excitation, isotopic and chemical abundance change across disks. Such surveys have the prospect to link Milky Way cloud studies to the evolution of galaxies, and provide key insights into the buildup of galaxies across the universe.

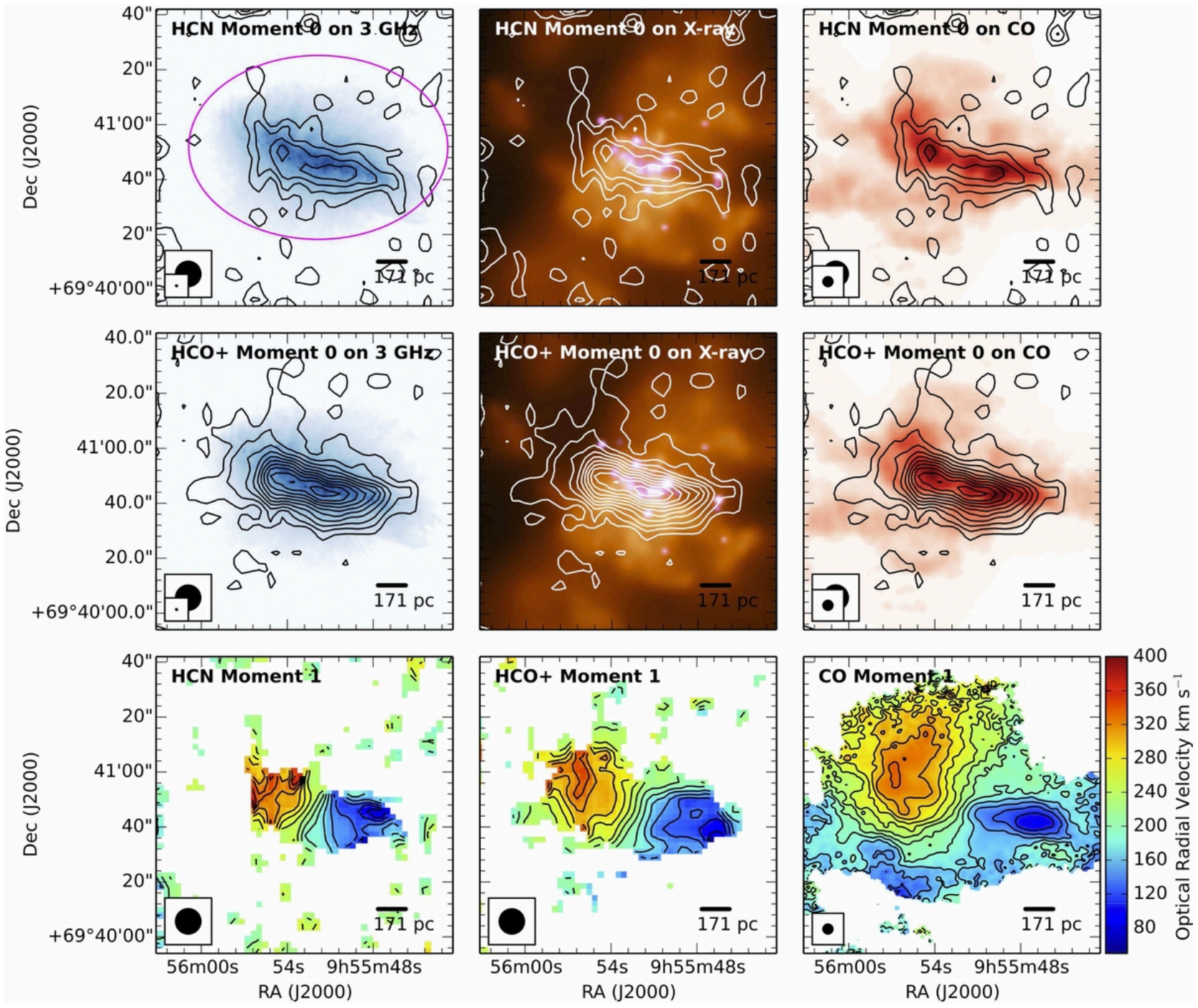

**Figure 3**. *GBT mapping of the prototypical starburst M82 in HCO⁺ and HCN using the double-beam W-band receiver reveals the extended structure of the dense gas (Kepley et al. 2014).*

These kinds of studies have already been begun with the two beam "W Band" receiver (Fig. 3). They will certainly accelerate using the newly deployed 16-beam ARGUS array, which renders the GBT competitive with ALMA to carry out such experiments. If large array receivers and backends capable of processing many lines become available in the future, the GBT could carry out cloud scale surveys of the local group and survey the physical state of the gas in hundreds of galaxies. This offers the prospect to build up fundamental knowledge about how cold gas connects to its host galaxies. For example: how does the amount of dense and very dense gas vary across galaxies? What drives this variation? How do chemical abundance patterns change across galaxies? Isotopic ratios? How widespread are shocks? How does emission break down between photon-, X-ray, and cosmic-ray dominated environments? And how does the local star formation rate depend on the physical state of the gas across galaxies?

The key capabilities here are the ~7.5" resolution of the GBT for the key lines at ~100 GHz, the excellent surface brightness sensitivity of a single dish telescope, and the mapping speed of array receivers. This allows the prospect to carry out detailed, multi-line spectroscopy across the nearby galaxy population. The angular resolving power of the GBT at ~100 GHz represents about an order of magnitude gain in beam area over the current best single dish telescopes. With current and next generation array receivers, the effective collecting area of the GBT (which is very large even accounting for the aperture efficiency) can be deployed simultaneously across many beams.

This makes the facility ideal to map resolved (~ few arcminutes) galaxies. ARGUS should reach a few mK per 25 km/s channel in ~4 hours over a 2x2 arcminute area, enough to access dense gas and PDR tracers like HCN, HNC, CN, $HCO^+$, and CS. Next generation receivers and backends will allow long integrations covering large bandwidth and area simultaneously, allowing one to push to lines 3-10 times fainter and so target shock tracers, optically thin isotopologues, and other important lines currently restricted to Galactic and starburst galaxy studies.

ALMA can do some of this science, but mosaicing will remain costly in terms of time and it requires spending time surveying at 3mm, which is not the best use of such an expensive facility. With a very large next-generation array receiver and backend, the GBT would outstrip ALMA's survey speed by a large factor. Even with instrumentation in commissioning (ARGUS), the GBT is already competitive with ALMA for mapping galaxies in dense gas tracers at modest resolution. But the mature status of the GBT allows the prospect to imagine spending hundreds or thousands of hours on such projects, while this remains farfetched for ALMA.

## 1.c. The Physics of Galaxy Clusters: the GBT as a Cluster Structure Machine

1. **What is the internal structure of Galaxy Clusters?**
2. **How frequently do we see evidence for accretion or feedback on cluster scales?**
3. **What is the mass distribution of clusters in the universe?**

Galaxy clusters are among the largest gravitationally-bound objects in the Universe, and their collapse is driven by the strongest fluctuations in the primordial matter power spectrum. As direct tracers of fluctuations in the early Universe, clusters are important signposts of the large-scale spatial distribution of dark matter, and therefore provide one of the most sensitive probes of the equation of state of dark energy (Mantz et al. 2015). Constraints relying on clusters are independent from, and complementary to, other cosmology experiments involving e.g. the cosmic microwave background (CMB), supernovae, and Baryon Acoustic Oscillations (BAO). Cluster cosmology includes: the growth of structure test, the dark matter power spectrum, the initial conditions of structure formation, and direct constraints on their angular and luminosity distances (Sasaki 1996; Pen, Seljak & Turok 1997; Carlstrom, Holder & Reese 2002; Allen, Evrard & Mantz 2011).

The past two decades have revolutionized our understanding of galaxy clusters. Once thought of as simple systems forming from isolated spherical collapse and in hydrostatic equilibrium, galaxy clusters have proved to be rich laboratories for astrophysical phenomena such as shocks and cold fronts, sloshing of the gas within a cluster's dark matter potential (Markevitch & Vikhlinin 2007; Blanton 2011), heating and the X-ray cavities/radio lobes produced by AGN outbursts (McNamara & Nulsen 2007; Blanton et al. 2010), and clumping in the outskirts due to subclusters infalling along the filamentary structure that connects clusters (Simionescu et al. 2011; Nagai & Lau 2011), all of which can impact the use of clusters as cosmological probes. Most of these phenomena were discovered through X-ray observations, and particularly through a significant investment in the X-ray facilities *Chandra* and *XMM-Newton*, both of which launched in 1999. Due to the strong cosmological dependence of X-ray surface brightness, X-ray cluster studies have mainly been limited to low redshift samples ($z<0.3$). With *Athena* – the X-ray successor to *Chandra* and *XMM-Newton* – scheduled to launch in 2028, the best way to advance our understanding of cluster astrophysics and evolution right now is to develop new observational tools to complement our existing (and aging) X-ray facilities.

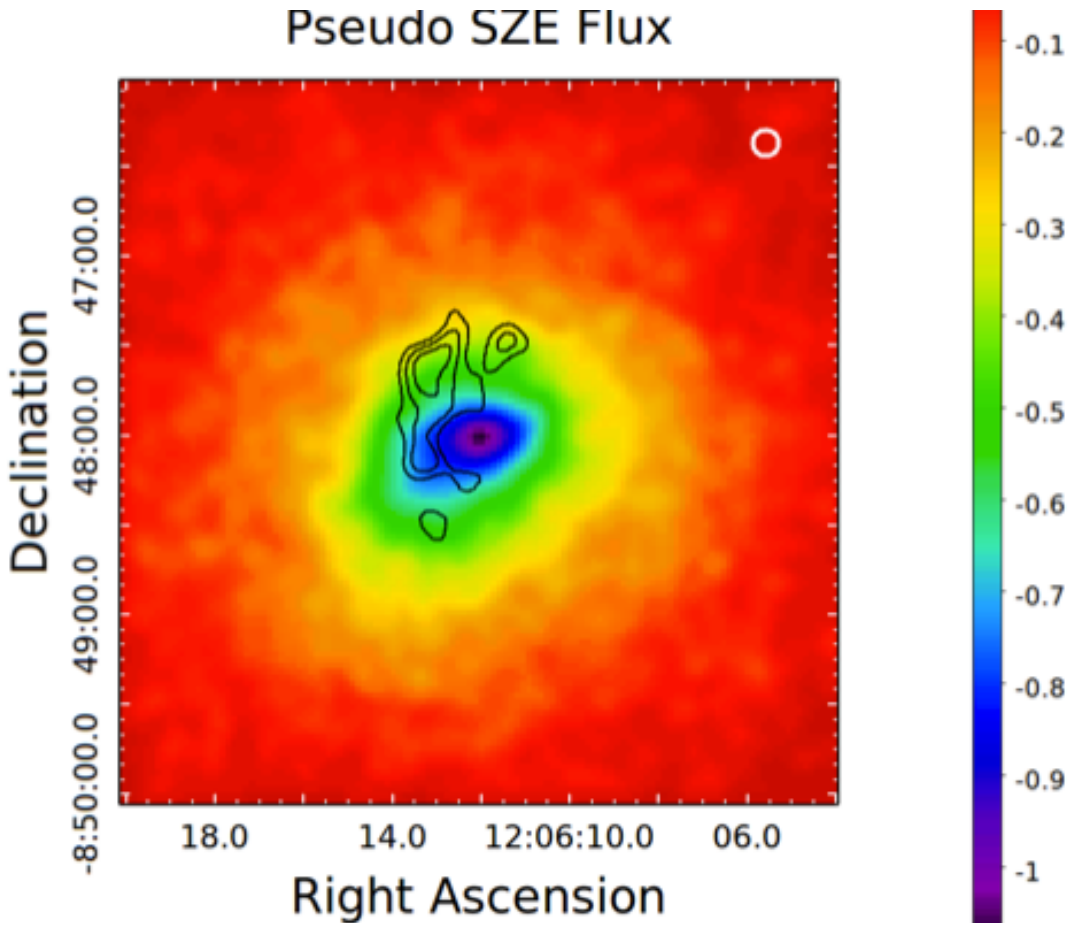

**Figure 4**. *MUSTANG-1 observations of the galaxy cluster MACS 1206-0847 reveal high-pressure structures invisible in X-rays, perhaps evidence for gas infalling into the cluster (Young et al. 2015).*

Fortunately, the mm waveband provides a powerful tool for cluster detection and astrophysical studies through the Sunyaev-Zeldovich (SZ) effect (Sunyaev & Zeldovich 1972). While other observational methods suffer from cosmological dimming (i.e. farther objects are less bright, as is the case at optical and X-ray wavelengths), the SZ effect is uniquely redshift-independent due to its nature as a fractional scattering of the CMB. This positions the SZ effect as particularly useful in the study of dynamically young, high-*z* clusters, as has been demonstrated in recent MUSTANG-1 and GISMO-1 results (Fig. 4). Further, the thermal SZ effect is directly proportional to the line-of-sight integral of thermal electron pressure ($\Delta I_{SZE} \propto n_e\, k_B T_e\, dl$), and its area integral therefore tracks gravitational energy. To the extent that clusters are in thermal pressure support, the integrated thermal SZ signal is a good proxy for mass (e.g. Marrone et al. 2011; Czakon et al. 2015), and SZ surveys are now yielding nearly mass-limited, *z*-independent cluster samples ideal for cluster cosmological studies.

ACT (Dunkley et al. 2011, Swetz et al. 2011), SPT (Plagge et al. 2010; Carlstrom et al. 2011), and *Planck* (Planck Collaboration et al. 2014) have each assembled large SZ cluster catalogs and have used these to place constraints on cosmological parameters such as the equation of state of dark energy and $\sigma_8$. However, SZ and X-ray cluster cosmological results are already systematics-limited, since the cosmological interpretation of their counts relies on the accurate determination of the scaling between the observable properties and cluster total mass (McCarthy et al. 2003); both X-ray and SZ depend sensitively on the level of non-thermal pressure support in clusters (e.g. turbulence, relativistic particles, and magnetic fields). The impact of systematics on the mass determination and cluster cosmology was recently poignantly demonstrated by the disagreement between the cosmological results from *Planck*'s SZ measurements (Planck Collaboration et al. 2014, XXIX) and those from *Planck*'s primary CMB measurements (Planck Collaboration et al. 2016, XVI). Because *Planck*'s 7–10′ beam does not resolve any but the nearest of clusters, SZ results from *Planck* rely on the 'Universal Pressure Profile' (UPP) of Arnaud et al. 2010 as a matched filter, which often categorizes multi-halo, merging systems as singular virialized systems. For example, some of the most spectacular $z > 0.5$ merging clusters imaged by GISMO and MUSTANG, were each entirely unresolved by *Planck*.

Recent simulations including more sophisticated AGN feedback and star formation physics appear to ease the tension between SZ and primary CMB results (Battaglia et al. 2013). However, it is important to test their predictions for the evolution of the pressure distribution, clumping, and overall gas fraction (Battaglia et al. 2015) through observations. This requires deep sub-arcminute resolution SZ observations of high-*z* clusters. Thus only a large format bolometer camera on the GBT, such as MUSTANG-2, can truly test the predictions for the SZ signal in detail, as the detection limits of even the ACT and SPT surveys are $3 \times 10^{14}$ $M_{\odot}$, and at > 1′ resolution, are not well suited to resolving astrophysical discrepancies or probing the SZ contributions from low mass systems (Hasselfield et al. 2013; Bleem et al. 2015). While blind SZ surveys have successfully discovered many previously-unknown clusters, such low-resolution, low-significance ($S/N \sim 4.5$) detections yield little information about the cluster kinetic and thermodynamical states. MUSTANG-2 SZ observations will probe the role and influence of the central AGN on the intracluster medium and how the pressure environment impacts the star formation rate (SFR) in galaxy clusters (where roughly half of galaxies at $z < 1$ reside).

## 1.d. The Origins of the Solar System: Planetary High-Frequency Science with the GBT

1. **How do isotope ratios compare across solar system objects, and how do they correlate with formation conditions in the proto-solar nebula?**
2. **How do solar system objects differ in gas production, dynamics, composition, and thermal structures?**
3. **What is the role of the solar wind and magnetic field in solar system interactions?**
4. **What lessons on star and planet formation can we learn locally by studying the current state and formation conditions of our own system?**
5. **How do exoplanetary systems compare in characteristics and evolution?**

Radio astronomy observations of solar system objects make unique contributions in planetary science. Continuum observations constrain thermal emission and the structure of dust and solid surfaces. Spectroscopic observations investigate chemical and dynamical properties of planetary and cometary atmospheres. *The instantaneous sensitivity of a large single-dish telescope is essential for solar system observations, because the targets have low flux, and dynamical, chemical, and electromagnetic environment changes can happen on rapid timescales.* A large instantaneous footprint on the sky is advantageous for highly extended, tenuous atmospheres, and for rapidly-moving nearby objects.

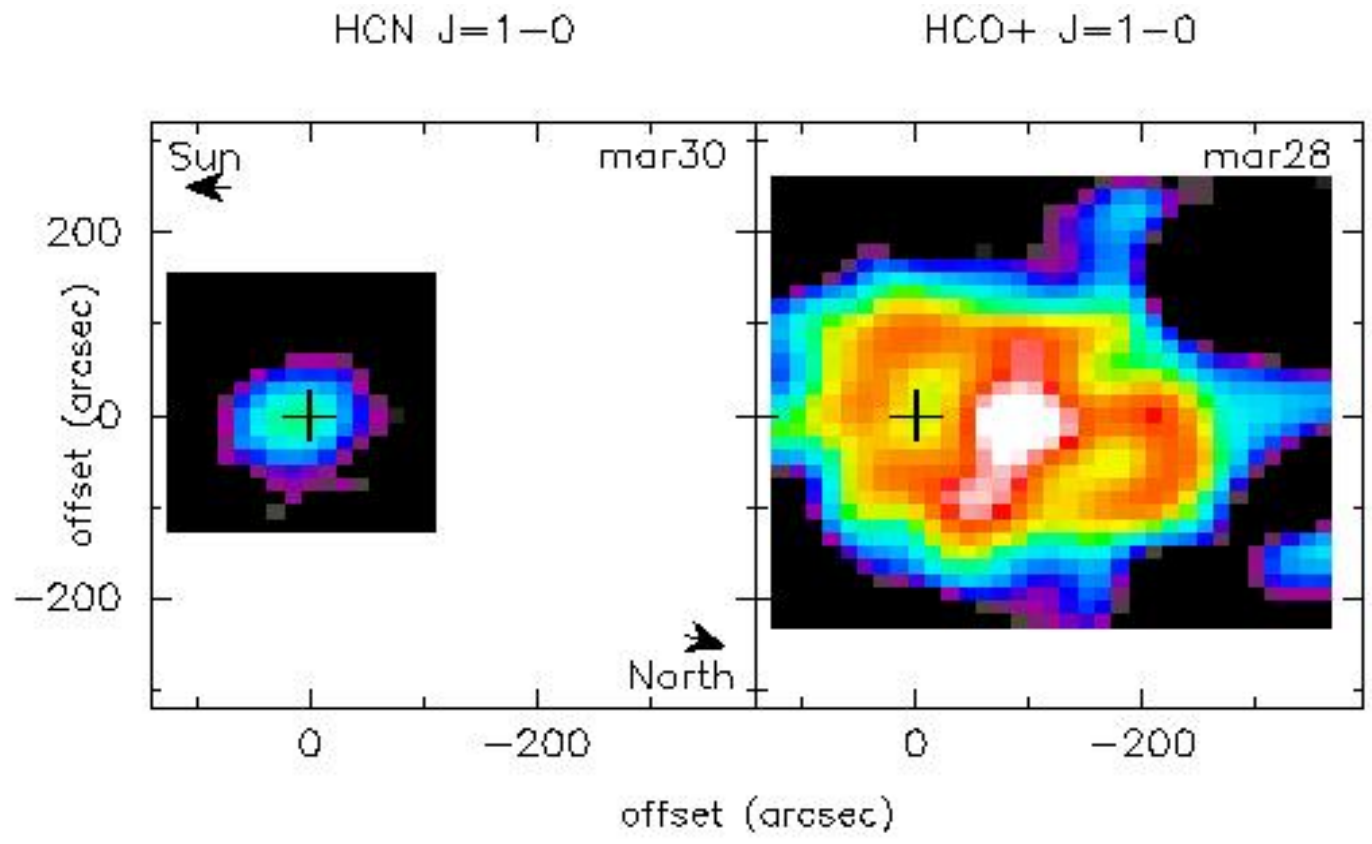

**Figure 5**. *Molecular maps for comet C/1995 O1 Hale-Bopp in March 1997 (Five College Radio Astronomy Observatory, from Lovell et al. 1998) of HCN (left) and HCO+ (right) at 90 GHz. While the extent of HCN was less than 200" on the sky, HCO+ extended in the anti-sunward direction by more than 300". The cross represents the ephemeris position of the comet nucleus.*

The GBT is a powerful tool for assessing important chemical species, isotope ratios, and chemical mixing ratios, molecular line observations of gases in the solar system (Fig. 5). Molecular column densities can be related through modeling to sublimation rates of ices, and line profiles offer an important constraint on production asymmetries, gas dynamics, and solar wind interactions. For Earth-approaching objects on eccentric orbits, the highest spatial resolution is obtained when they are at close approaches, which often occurs at low solar elongations inaccessible to instruments at other wavelengths. Thus, the GBT observations provide a means of extending the temporal baseline and filling gaps in multi-wavelength observational campaigns. Radio spectroscopy provides line-of-sight velocities for expanding gases in cometary atmospheres. In combination with optical and IR imaging, these can be used to produce a full three-dimensional dynamical picture of the gas.

High spectral resolution 90 GHz spectroscopy uniquely identifies spectral lines and separates overlapping and hyperfine line structures that are often blended in optical/IR spectra. Chemical mixing ratios are often hard to constrain when column densities and gas production rates are estimated by vastly different techniques at different wavelengths. Wide-band, high-resolution mm spectroscopy of solar system atmospheres offers the opportunity to constrain mixing ratios for related species in a self-consistent manner. Of particular interest to cometary chemistry are the relationships between parent species, directly sublimated from the nuclear ice, and daughter species that arise after photo-dissociation of those molecules in the solar system UV environment.

Molecular mapping observations also probe the radial profile of the gas density in comets: these profiles extend measurements the nucleus made at other wavelengths. The combination probes of gas acceleration and asymmetries and to constrains gas excitation or non-nuclear sources of gas. Questions persist about HCN and CN branching ratios that would be addressed with high-sensitivity near-simultaneous spectroscopy of both species in the 90GHz band. Important differences between neutral and ionized species as well as different isotopes and isomers shed light on solar wind interactions, ion-molecule chemistry, and the formation of the solar system in different heliocentric distance regions. CO and related species are likely to be incredibly important drivers for gas production in the outer solar system, where water sublimation is suppressed. Using the GBT to extend observations to greater heliocentric distances, and to follow comets on their outbound post-perihelion journeys, provides an important contribution in enabling comparison of the inbound and outbound behavior of sublimation from comets, chemical mixing ratios of different gas species, and isotope ratio evolution.

Spectroscopic mapping is critical for ionized species, which respond rapidly to variations in the solar wind. Ionized species, such as $HCO^+$, form a much larger coma with greater anti-sunward extension (Figure 5). Observing this requires the instantaneous sensitivity and wide field of view of the GBT to characterize the spectral lines on appropriate (minutes to hours) timescales. The GBT also allows the option of monitoring observations to follow a comet over many weeks and months, as it crosses the solar equator and higher latitudes. Interferometric observations may be compromised by a comet's rapid motions across the *u*-*v* plane, or require multi-instrument coordination to provide zero-spacing observations to avoid flux loss for faint, extended objects, are poorly suited to these observations, while the GBT configuration is ideal. The same arguments can be applied to planetary atmospheres, and potentially to exoplanet atmospheres and exoplanet-stellar interactions as those fields develop. In addition to spectroscopy and a focus on molecules in solar system atmospheres, GBT has potential for contributions to scientific development in dust and solid surfaces as well.

Radio observations in the solar system are well-suited to the study of cool solid surfaces and low-temperature dust. Little is known about the dust size distribution outside the visible and IR bands. High-frequency radio observations with the sensitivity of GBT have the potential for characterizing the dust temperature and size distributions for comets and other solar system dust sources.

Thermal emissions in these energy ranges generally arise from below the surfaces of the bodies, offering a unique opportunity to constrain the composition and structure of planets and small solar system bodies using mm-wave observations. Such constraints are a critical supplement to in situ spacecraft observations, which are rare and costly. For potentially hazardous asteroids, thermal observations can help constrain the near-surface density, porosity, composition, and cohesion. This could become important should such an object have the potential of a collision with Earth. For main-belt asteroids, planets, and planetary satellites, thermal observations constrain the history, composition, and thermophysical properties of these bodies, helping to tie them to models of solar system formation and evolution.

Both passive radio astronomy and radar ranging of planets, moons, asteroids, comets, rings, and dust have been carried out with success at lower frequencies, and the GBT has played an important role in receiving bi-static radar observations from Arecibo, Goldstone, and other radar facilties. As millimeter-wave technology for radar is developed and implemented at other observatories, the GBT will continue to be important as a receiver for transmissions from radar stations. In some circumstances, the bi-static configuration is necessary because the round-trip time from transmit to receive is too short for operation at a single facility. In other circumstances, the bi-static configuration adds to the science value in observing geometry, reception time, and sensitivity, which is increasingly important as radar returns diminish with the fourth power of the object's distance. Radar transmissions at high frequencies, received at GBT, have applications for near-Earth and potentially hazardous asteroids, meteor streams, and even for space junk. Radar observations help refine orbital parameters, assess spin and orbital evolution and relate those to dynamical factors and causes. This is important in an era where spin and orbital evolution of small bodies is understood to be critical for assessing impact hazards.

## 1.e. The GBT as a Key Element of the Global mmVLBI Network

1. **What are the physics and masses of supermassive black holes?**
2. **What determines the structure of the inner regions of relativistic jets in AGNs?**

The only current technique capable of viewing directly the subparsec-scale regions of jets in active galactic nuclei (AGNs) is imaging with very long baseline interferometry (VLBI). VLBI has produced major breakthroughs in our understanding of the IR, optical, X-ray, and gamma-ray emission, where techniques to study very small scales are necessarily more indirect. For example, the discovery of apparent superluminal motions in blazars led to the standard model of relativistic jets that are responsible for essentially all of the nonthermal emission in the nucleus.

With the commissioning of the ALMA phasing capabilities, mm-wave VLBI has gained an extremely powerful southern station. By far the most sensitive baselines in the 3mm VLBI network are those involving ALMA in the south, Effelsberg in the east, and the GBT in the west. For southern sources, such as the Galactic Center, the common visibility of ALMA and the GBT (which have similar longitudes) makes this baseline a crucial element of the 3mm VLBI network.

Global VLBI provides angular resolution of about 70 microarcseconds at a wavelength of 3 mm. This allows us to view AGN jets quite close to the accreting supermassive black hole that powers the AGN. In the case of M87, mm-VLBI is able to resolve the jet to within about 15 Schwarzschild radii of the black hole (Hada et al. 2011).

In order to take advantage of this extraordinary resolving power, the VLBI array needs to include antennas with very high sensitivity located at crucial positions in the array. The GBT provides such high sensitivity on the northern side, while the LMT can do so closer to the equator, the IRAM 30-meter across the Atlantic, and ALMA on the southern end. Together, these large mm-wave antennas provide Earth-diameter baselines with sufficient sensitivity to overcome issues related to weak sources, extended (larger than a few tenths of a milliarcsecond) source structure that is partially "resolved out," atmospheric attenuation, and phase instability. *There is no other large mm-wave antenna to take the GBT's place in the northern US, and without the GBT, mm-wave image quality would suffer, reducing the dynamic range and fidelity of mm-wave VLBI.*

Main scientific applications of mm-wave VLBI in the next years include studies of the black holes in the Galactic Center and in M87 to probe the strong gravity regime, mapping masers in star-forming regions, evolved carbon stars, and in the disks around AGN. The velocity structure of masers in the nuclei of AGNs can be used to measure black-hole masses or even infer direct distances. Mm-wave VLBI also remains a main tool to understand the physics behind jets and active galactic nuclei (AGN). It allows studies of the regions of relativistic jets where the flow is accelerated up to near-light speeds and collimated to within two degrees or less. It can be used to relate the fine mm-wave structure of AGN jets to the optical and gamma-ray emitting regions. And mm-wave VLBI can be used to map the magnetic field structure near the base of AGN jets or in young stellar objects. Faraday rotation maps can also probe the interstellar medium in the nuclei of AGNs.

# 2. Operations and Instrumentation of a High Frequency GBT

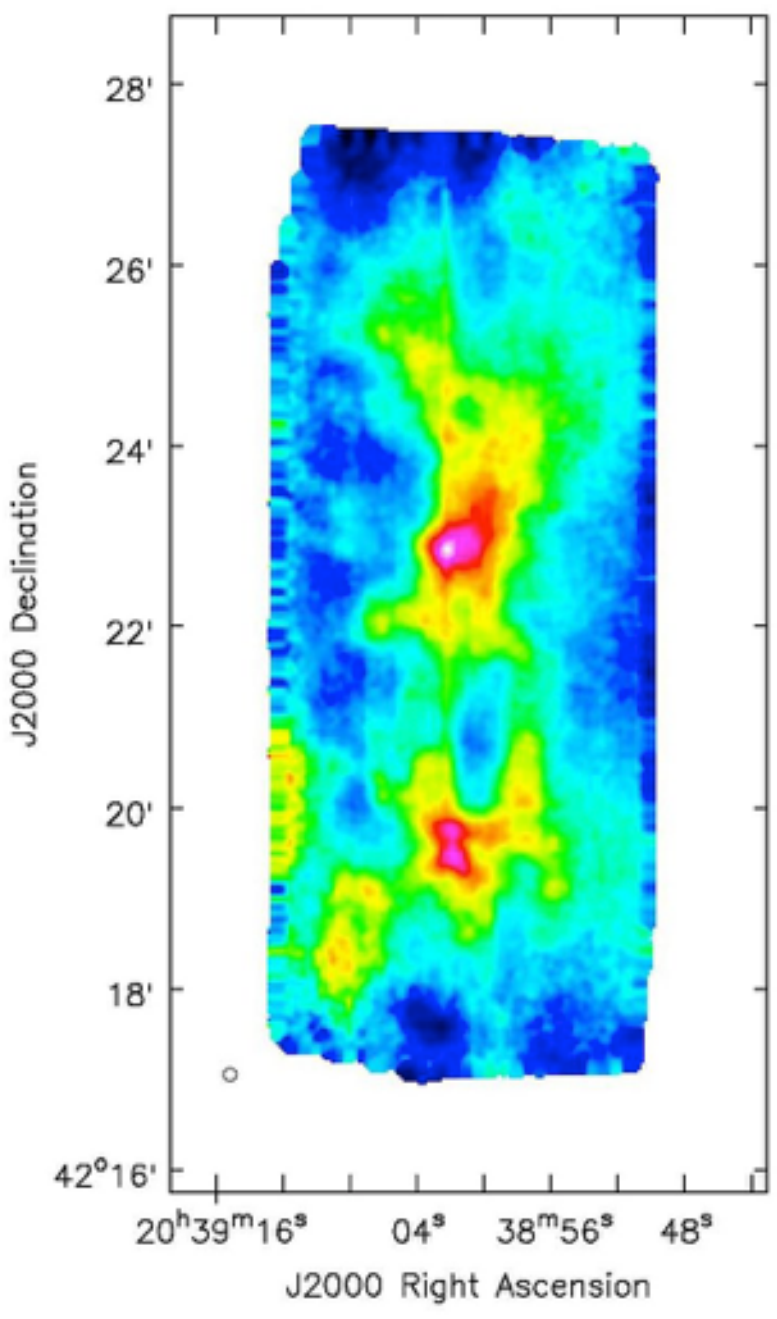

**Figure 6**. *A GBT test map of $^{13}CO$ in DR21 obtained with the 16-pixel array ARGUS in September 2016.*

**Weather:** Weather at the GBT compares favorably to previous productive mm-wave sites such as the FCRAO 14-m antenna. At an altitude of 844 m (2770 ft), and with winter temperatures routinely well below 0 C, atmospheric water vapor can be very low for a continental site. Precipitable water vapor (PWV) at the GBT site is below 10 mm about 50% of the year. The mode in the distribution of PWV is about 5-6 mm, with a long tail of high water in the summer and lower water vapor in the winter. For reference, OVRO considered PWV around 5 mm the threshold for 1 mm observations, while these are average winter conditions for the IRAM 30-m telescope. Observations at 3mm are common across observatories for PWV between 5 and 10 mm.

The small beam size of the GBT imposes another constraint. Pointing errors become comparable to the size of the GBT beam at high frequencies for wind speeds above about 3 m/s (6.7 mph), degrading high frequency observations. Experience with the site shows that the time available when both water vapor and wind are low amounts to *80 to 150 12-hour nights per year, roughly 1,000 to 2,000 hours*.

This available time could be increased by expanding high frequency observations into the day time. High frequencies are currently only done at nighttime to avoid spending an unreasonable amount of time on surface updates in response to changing solar illumination. GBT scientists are exploring options that allow very rapid surface measurement and correction, for example high-precision optical mapping of the dish surface. These methods offer the prospect of expanding the time available through routine daytime high frequency observations for a modest investment.

**Prospects for Instrumentation:** The GBT already has impressive high-frequency heterodyne mapping capabilities, with a 7-element focal plane array (FPA) at K-band, a 16-element FPA in the 3-4 mm band, and a 64-element experimental phased array also at mm-waves. Additionally, the 223-element MUSTANG-2 bolometer camera allows for continuum 90 GHz observations. As emphasized above, placing these arrays on the GBT takes advantage of the high resolution afforded by it, while achieving the unparalleled surface brightness sensitivity of a single dish. Figure 6 illustrates this capability, showing $^{13}$CO $J$ = 1-0 line emission from the DR21 region. This map covers about 2' by 10' on the sky, with the beam size shown in the lower left corner. It was obtained in only ~3 hours of commissioning time using the 16-pixel ARGUS array.

We are entering the era when even larger heterodyne imaging arrays are possible. Moving forward, it will be possible to increase dramatically the number of pixels, to explore better coupling/sampling for large-scale mapping arrays, and to implement flexible digital backends that can make real-time trade offs between the numbers of pixels used and the bandwidth per pixel (e.g., allowing use of many pixels for to map a single line or fewer pixels with much broader bandwidth to carry out velocity-resolved line surveys in chemically complex regions).

These goals are achievable with modest cost and in the near future. ARGUS, for instance, is designed as a scalable prototype, with highly-integrated receiver/mixer modules that plug into carriers. Straightforward extension of the ARGUS technology could yield a 10×10 array, yielding impressive mapping speeds (Table 1). Meanwhile, PHAROH, an experiment in millimeter-wave phased arrays, could solve the sample-spacing problem imposed by diffraction-limited optics, allowing highly efficient mapping of smaller extended regions. The main limit on the size of both focal plane arrays and phased arrays is signal transport from the arrays to appropriately shielded digital processing, and to some extent the digital processing itself. Neither of these are fundamental problems.

These developments offer a path toward the ultimate, eminently achievable goal of capturing all of the photons across the GBT's field of view, up to 900 independent pixels in the 3-4 mm band. Such next-generation arrays will massively increase the scientific throughput of the GBT, providing a very high scientific return for the investment in the telescope, and leading to advances well beyond those possible with the current generation of arrays. Even greater gains can be made in the future by increasing the telescope efficiency with a robust active surface control system.

**University-Observatory-NSF Collaborations:** MUSTANG-2, ARGUS, and PHAROH were built by collaborations between university groups and the GBT with support from the NSF. This kind of university-NSF-observatory partnership is enabled by the existence of a large single-dish telescope that can produce excellent science given cutting edge instrumentation. Continued support and development for GBT high frequency capabilities can be expected to enable more such partnerships, spurring the development of ever more powerful radio cameras and helping to ensure the health of university-based instrumentation groups while training the next generation of mm-wave instrumentalists.

---

[1] University of Colorado, Boulder, CO, USA
[2] California Institute of Technology, Pasadena, CA, USA
[3] University of Maryland, College Park, MD, USA
[4] University of Texas, Austin, TX, USA
[5] Stanford University, Palo Alto, CA, USA
[6] NRC Herzberg Institute of Astronomy and Astrophysics, Victoria, BC, Canada
[7] NASA Jet Propulsion Laboratory, Pasadena, CA, USA
[8] Harvard University, Cambridge, MA, USA
[9] Boston University, Boston, MA, USA
[10] The University of Newcastle, Callaghan, NSW, Australia
[11] Ohio State University, Columbus, OH, USA
[12] Green Bank Observatory, Green Bank, WV, USA
[13] Agnes Scott College, Decatur, GA, USA
[14] University of Arizona, Tucson, AZ, USA
[15] National Radio Astronomy Observatory, Charlottesville, VA, USA
[16] European Southern Observatory, Garching-bei-München, Germany
[17] University of Massachusetts, Amherst, MA, USA